\documentclass[journal]{IEEEtran}
\usepackage{cite}
\usepackage{times,amsmath,amssymb,psfrag}
\usepackage{graphicx}
\usepackage{url}
\usepackage{bm}
\usepackage{multicol}
\graphicspath{{./pics/}} \DeclareGraphicsExtensions{.pdf,.jpg,.png}
\usepackage{mdframed}
\usepackage{xcolor}
\usepackage{hyperref}
\usepackage{epsfig}
\usepackage{float}
\usepackage{array}
\usepackage{multirow}
\usepackage{etoolbox}
\usepackage{nomencl}
\usepackage{algorithm}
\usepackage{algorithmicx}
\usepackage{algpseudocode}
\usepackage{amsmath}

\renewcommand\nomgroup[1]{%
  \item[\normalsize\itshape\bfseries
  \ifstrequal{#1}{I}{Parameters/States of Composite Load Model}{%
  \ifstrequal{#1}{P}{Notation for Algorithm}{%
  \ifstrequal{#1}{N}{Notation1 for Algorithm}{%
  \ifstrequal{#1}{X}{Other Symbols}{}}}}]%
  }
\makenomenclature
\begin{document}

\title{Data-Driven Based Method for Power System Time-Varying Composite Load Modeling}
%
%
%

\author{Jian~Xie,~\IEEEmembership{Student Member,~IEEE,}
	Zixiao~Ma,~\IEEEmembership{Student Member,~IEEE,}
	Shanshan~Ma,~\IEEEmembership{Student Member,~IEEE,}
	and~Zhaoyu~Wang,~\IEEEmembership{Member,~IEEE}
	\thanks{The authors are with the Department of Electrical and Computer Engineering, Iowa State University, Ames,
		IA, 50011 USA e-mail: (jianx@iastate.edu;zma@iastate.edu;sma@iastate.edu;wzy@iastate.edu).}
	
}
%
%

\markboth{IEEE TRANSACTIONS ON SMART GRID,~Vol.~14, No.~8, JANUARY~2019}%
{Shell \MakeLowercase{\textit{et al.}}: Bare Demo of IEEEtran.cls for IEEE Journals}
%



\maketitle

\begin{abstract}
	Fast and accurate load parameters identification has great impact on the power systems operation and stability analysis. This paper proposes a novel transfer reinforcement learning based method to identify composite ZIP and induction motor (IM) load models. An imitation learning process is firstly introduced to improve the exploitation and exploration process. The transfer learning process is then employed to overcome the challenge of time consuming optimization when dealing with new tasks. An Associative memory is designed to realize demension reduction and knowledge learning and transfer between different optimization tasks. Agents can exploit the optimal knowledge from source tasks to accelerate search rate and improve solution accuracy. The greedy rule is adopted to balance global search and local search. Convergency analysis shows that the proposed method can converge to the global optimal solution with probability 1. The performance of the proposed ITQ appraoch have been validated on 68-bus system.  Simulation results in multi-test cases verify that the proposed method  has superior convergence rate and stability.
\end{abstract}


%
\IEEEpeerreviewmaketitle

\section{Introduction}
%
%
%
%
\IEEEPARstart{A}{s} an important part of power system, electrical load model has a great impact on the stable operation of power system \cite{Arif2018,Bai2009,Wang2019} Incorrect load models may lead to a completely contrary  results for the system operation status and stability evaluation\cite{Ma2006,Knyazkin2004,Han2009,Ma2008}. Owing to the characteristics of random time variance, complex composition and nonlinearity, fast and accurate load modeling still remains the most challenging problem.
Therefore, it is imperative to achieve load model parameters accurately and fast to help provide more reliable results for power system opreation.

Based on load models' characteristics, conventional load models can be categorized as three types : static load models, dynamic load models and composite load models. For static load models, active and reactive power can be expressed as the functions of bus voltage and frequency. Common static load models include constant impedance-current-power (ZIP) model\cite{Kundur1994}, exponential model\cite{Emodel} and frequency dependent model\cite{Fremodel}. Dynamic load models can present the relationship between load powers (i.e. active power and reactive power) and bus voltage and time. Representative dynamic loads are induction motor (IM) load and exponential recovery load model (ERL)\cite{ExDynamicM1}. IM load model is considered as a physically-based model since it is derived from the equivalent circuit of an induction motor. The ERL load
model is usually utilized for describing loads that recovery slowly. Numerous research shows that single static model or dynamic model cannot sufficiently replicate the dynamic behavior of the actual load. Therefore, a composite load model combining ZIP model with the IM model has been widely adopted by most of the utilities to represent the actual load models since it aggregate the static and dynamic properties, which can provide a more accurate characteristics\cite{Ma2006,Han2009}.

Currently, most research foucses on measurement based load identification and parameters estimation.  Measurement based methods can be classified into two categories: artificial neural network based methods  and optimization based methods. The artificial neural network (ANN) based methods do not require the pre-defined physical load models and it can update load outputs (i.e. active and reactive powers of loads) based on the measurements quickly. In \cite{Miranian2013}, a local model networks (LMN) based method was proposed to identify the load models, and it shows a fast training rate and convergence rate. A deep learning based  technique was proposed in \cite{Cui2019} to identify time varying load parameters.

Optimization based parameter estimation algorithms usually pre-define a load structure and then try to search the optimal parameters to minimize the error between the actual power measurements and the estimated power response. These methods can be divided into statistical techniques and heuristic techniques based methods. Common statistical technique based search methods include least square (LS) method, maximum likelihood method and gradient-based method. A least square based method was proposed in \cite{Liu2002} to minimize the difference between estimated load powers and true load powers. \cite{Hiskens2001} utilized a weighted LS method to estimate the parameters of a first order induction motor. These least squares methods are sensitive to the effects of outliers. Also, it is difficult to determine the exact weight when the weights are estimated from small numbers of replicated observations. A maximum likelihood  approach was adopted in \cite{KAMOUN1992} to estimate the load parameters. Two disadvantages of this method are that it requires strong assumptions on the  data structure and maximum likelihood is sensitive to the choice of starting values. In addition, it is very CPU-consuming and thus extremely slow.  In \cite{DeKock1994}, a gradient-based method was proposed to estimate parameters of a fifth-order induction motor load. It is very sensitive to the learning rate and depends on proper initialization. Sometimes it could lead to erroneous results without proper learning rate and initial values. 
As for heuristic techniques based methods, simulated annealing algorithm, genetic algorithm (GA)\cite{Wen2003} have been widely adopted to estimate the parameters of load models. The simulated annealing (SA) technique does not require detailed mathematical descriptions of load models\cite{Knyazkin2004}. However, the runs often require a great deal of computer time, and many runs may be required to obtain good results. GA based method is sensitive to the initial population used. In addition, premature convergence is another issue that should be considered consider when solutions are generated. An improved particle swarm optimization method has been applied in \cite{Regulski2015} to identify the unknown composite load model parameters. It was proved that the IPSO based method has superior convergence properties compared to GA based method. Unfortunately, most of the above methods are unable to exploit the prior optimization knowledge when dealing with new optimization tasks, which will results in a low efficient search when dealing with new tasks.

In applications invovling the non-linear optimization problem, reinforcement learning (RL) methods can be adopted to obtain the optimal solution\cite{re2018}. During the reinforcement learning process, agents excute actions and update their states based on the exploration and explitation rules. As one of the common RL method, Q learning has been widely used for online optimization and control. Unfortunately, like heuristic approaches,  RL methods also suffer from unable to storing the prior knowledge and it is time consuming to identify a lager number of load parameters when dealing with new optimization tasks.
Aiming to exploiting the prior knowledge in the source tasks to deal with new tasks and can significantly accelerate the computation time, transfer learning has drawn remarkable interests in data mining and machine learning\cite{Pan2010}. Motivatied by this, a novel approach aggrates Q-learning, transfer learning and imitation learning\cite{Zhang2018} is proposed to mitigate the computational burden and improve the identification accuracy in load model parameters identification.  Druing the transfer learning process, the prior learned knoeledge can lead a RL agent to explore more effectively when dealing with new task. The imitation learning process is introduced to guide the RL agent to execute a more informative exploration instead of a random one.

The rest of the paper is structured as follows. Section II describes the composite load model structure. Section III presents the basic principles of transfer and imitation learning based Q-learning (TIQ). The framework of TIQ based load model parameters identification is given in section IV. Simulation results are presented in Section V and Section VI concludes this paper.  

\section{Composite load model structure}
\begin{figure}[htb]
	\centering                                   
	\includegraphics[width=8cm]{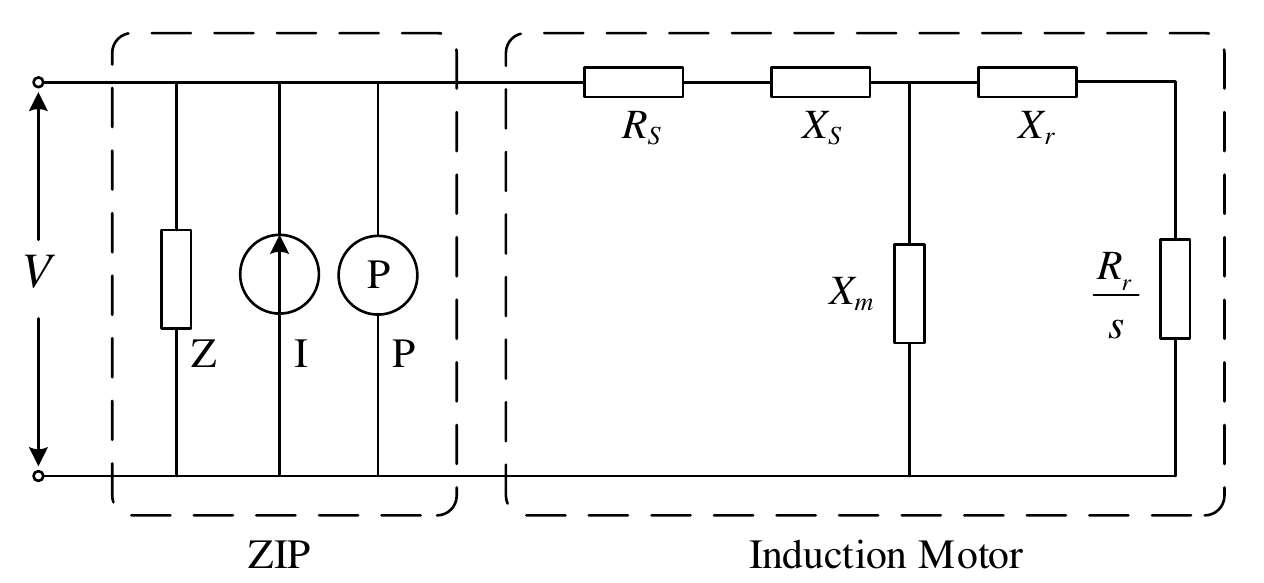}
	\caption{Equivalent circuit of this composite load model\cite{Wang2019}}
	\label{f1}
\end{figure}

As mentioned before, the composite load model consisting of ZIP and an induction motor and an compensator is analyzed in this paper. The composite load aggrated with ZIP and an IM is the most commonly used load model in industry. An equivalent circuit of this composite load model is shown in Fig. \ref{f1}. Considering the ZIP model and the IM model are mathematically decoupled, we can describe  these two models independently.

From the components in ZIP model, the mathematical description of the active and reactive power are expressed as follows:
\begin{eqnarray}
\label{Equation.ZIP_P}
\ {P_{ZIP}} = {P_{ZIP,0}}\left( {{a_{p}}{{\left( {\frac{{{V}}}{{{V_0}}}} \right)}^2} + {b_{p}}\left( {\frac{{{V}}}{{{V_0}}}} \right) + {c_{p}}} \right)
\end{eqnarray}
\begin{eqnarray}
\label{Equation.ZIP_Q}
\ {Q_{ZIP}} = {Q_{ZIP,0}}\left( {{a_{q}}{{\left( {\frac{{{V}}}{{{V_0}}}} \right)}^2} + {b_{q}}\left( {\frac{{{V}}}{{{V_0}}}} \right) + {c_{q}}} \right)
\end{eqnarray}
{where $P_{ZIP,0}$,  $Q_{ZIP,0}$ and $V_{0}$ are active, reactive power and root-mean-square (RMS) value of voltage in the steady state before disturbance, respectively.  In addition, $Q_{ZIP,0}$, $a_{p}$, $b_{p}$ and $c_{p}$ satisfy $a_{p}+b_{p}+c_{p}=1$, and $a_{q}$, $b_{q}$ and $c_{q}$ satisfy $a_{q}+b_{q}+c_{q}=1$.} 

With respect to the induction motor part shown in Fig. \ref{f1}, ${R_s}$ denotes the stator resistance, ${R_r}$ represents the rotor resistance, ${X_s}$ is the stator reactance, and ${X_r}$ is the rotor reactance, respectively. ${X_m}$ represents the magnetizing reactance, and ${s}$ is the slip. The induction motor model can be expressed as follows:
\begin{eqnarray}
\label{Equation.3}
\ \frac{{d{E'_{d}}}}{{dt}} =   -\frac{1}{T'}\left[ {E'_{d}} + \left(X-X'\right) I_{q}\right] - \left({\omega-1}\right){E'_{q}} 
\end{eqnarray}

\begin{eqnarray}
\label{Equation.4}
\ \frac{{d{E'_{q}}}}{{dt}} =   -\frac{1}{T'}\left[ {E'_{q}} - \left(X-X'\right) I_{d}\right] + \left({\omega-1}\right){E'_{d}} 
\end{eqnarray}

\begin{eqnarray}
\label{Equation.5}
\ \frac{{d{\omega}}}{{dt}} = -\frac{1}{{2{H}}}\left[ {T_{0}}\left({{A{\omega^2}+B{\omega}+C}}\right) - \left({E'_{d}}{I_{d}} +{E'_{q}}{I_{q}}\right) \right] 
\end{eqnarray}
{where $H$ is the rotor inertia constant; $A$, $B$ and $C$ denote the torque coefficient and satisfy $A+B+C=1$;$\omega=1-s$ represents the rotation speed of the induction motor;$E'_d$ and $E'_q$ refer to the d-axis and q-axis transient EMF of the induction motor.}
\[T'=\frac{X_r+X_m}{R_r},{X=X_{s}+X_{m}}\] \[{X'=X_s+\frac{X_{r}X_{m}}{X_r+X_m}}\] 

${I_d}$ and ${I_q}$ are the $d$ and $q$ axes currents, which can be expressed as:
\begin{eqnarray}
\label{Equation.I_D}
\ {I_{d}} = \frac{{{R_{s}}({U_{d}} - {E'_{d}}) + {X'_t}({U_{q}} - {E'_{q}})}}{{R_{s}^2 + {X'}^2}}
\end{eqnarray}
\begin{eqnarray}
\label{Equation.I_q}
\ {I_{q}} = \frac{{{R_{s}}({U_{q}} - {v'_{q}}) - {X'}({U_{d}} - {E'_{d}})}}{{R_{s}^2 + {X'}^2}}
\end{eqnarray}
{where the $d$-axis bus voltage $U_{d}$ and the $q$-axis bus voltage $U_{q}$ satisfy the following equation.}
\begin{eqnarray}
\label{Equation.Vt}
\ {U} = \sqrt {U_{d}^2 + U_{q}^2}
\end{eqnarray}

With the states, parameters and the bus voltage, the active and reactive power of the IM model can be expressed as

\begin{eqnarray}
\label{Equation.PIM}
\ \begin{array}{l}
{P_{IM}} = U_{d}I_{d}+U_{q}I_{q}
\end{array}
\end{eqnarray}
\begin{eqnarray}
\label{Equation.QIM}
\ \begin{array}{l}
{Q_{IM}} = U_{q}I_{d}-U_{d}I_{q}
\end{array}
\end{eqnarray}

With the ZIP and IM model aggregated, the total active and reactive power of the composite load model can be expressed as:

\begin{eqnarray}
\label{Equation.Ptotal}
\ {P} = {P_{ZIP}} + {P_{IM}}
\end{eqnarray}
\begin{eqnarray}
\label{Equation.Qtotal}
\ {Q} = {Q_{ZIP}} + {Q_{IM}}
\end{eqnarray}

For the composite load model, another important parameter to be identified is the initial active power proportion of the induction motor to the total load, which is defined as :
\begin{eqnarray}
\label{Kpm}
\ K_{pm}=\frac{P_{Im,0}}{P_0}
\end{eqnarray}
{where $P_0$ denotes the initial active power of the composite load before disturbance and $P_{Im,0}$ is the initial active power of the equivalent induction motor.}

Traditionally, the 13 parameters in equations (1)-(13) need to be estimated to identify a accurate load model, which are
\[\theta=\left[ {{R_{s}},{X_{s}},{X_{m}},{X_{r}},{R_{r}},{H},A,B,{a_{p}},{b_{p}},{a_{q}},{b_{q}},{K_{pm}}} \right]\]

The objective function for parameter identification is to ensure the differences between the estimation active(reactive) power and the measured active(reactive) power is minimum, which can be expressed as:
\begin{eqnarray}
\label{obj}
\min h{(\theta)}= \frac{\sum\limits_{k = 1}^L {\left[ {{{\left( {{P_\theta}\left( k \right) - {P}\left( k \right)} \right)}^2} + {{\left( {{Q_\theta}\left( k \right) - {Q}\left( k \right)} \right)}^2}} \right]}} {L}
\end{eqnarray}
{where $L$ is the total sampling points of the measurements. ${P_\theta}\left( k \right)$ and ${P}\left( k \right)$ are estimated and measured active power respectively; ${Q_\theta}\left( k \right)$ and ${Q}\left( k \right)$ are estimated and measured reactive power respectively.}

However, in real power systems, load components vary frequently and power system operation status could change dramatically under various fault conditions, which can result in different load parameters and re-estimation may consume a lot of time. Therefore, traditional estimation methods may not be able to identify the exactly time varying load models and lack generalization.

\section{Imitation and Transfer Q learning} 
\begin{figure}[htb]
	\centering                                   
	\includegraphics[width=8cm]{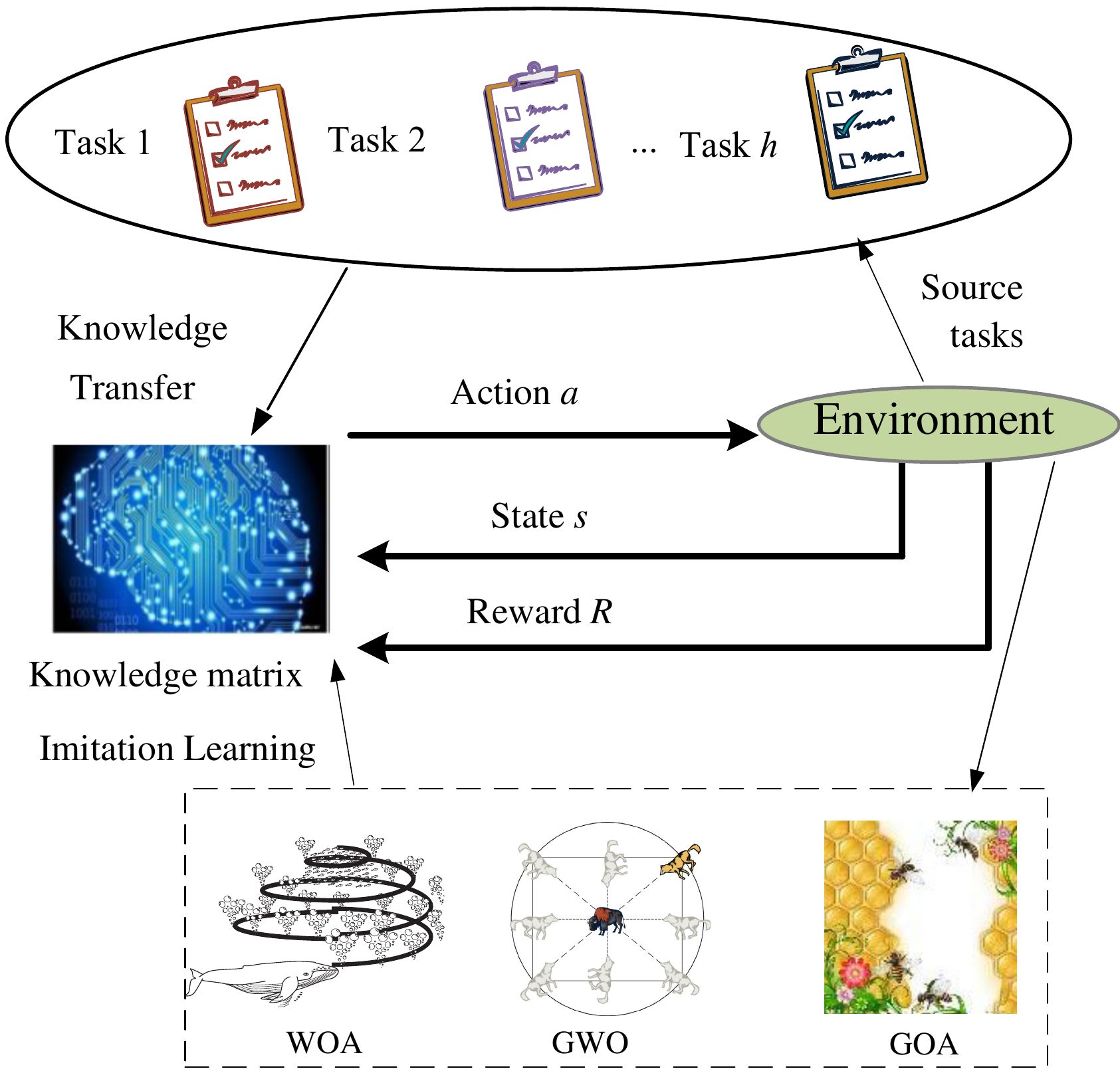}
	\caption{Overall scheme for ITQ}
	\label{f2}
\end{figure}

The overall process to implement the imitation and transfer Q learning is shown in Fig. \ref{f2}. There are 4 main steps: 1) Learning knowledge from the source tasks based on Q learning; 2) Accelerating learning rate and improving optimal solution via imitation learning; 3) Define and compute the similarities between source tasks and new task; 4) Dealing new task by transfer learning.

\subsection{Q learning  }
Like other classical RL methods, Q learning aims to obtain an optimal policy such that the reward is maximized. For Q learning algorithm, agent observes the current state and execute an action. The system observes the corresponding results and samples a reward to the agent. The agent recieves the reward and update the Q value stored in Q-tabel. Then, a new action is selected, resulting in a new state and a new reward. By  continually expolitation and exploration, agent will eventually obtain the optimal Q tabel and policy. However, there are two disadvantages for traditional Q learning method: 1) the dimension of knowledge matrix will increase dramatically if the number of controllable variables or the alternative actions increase; 2) a single RL agent leads to a low knowledge learning efficiency.

\begin{figure}[htb]
	\centering                                   
	\includegraphics[width=8.5cm]{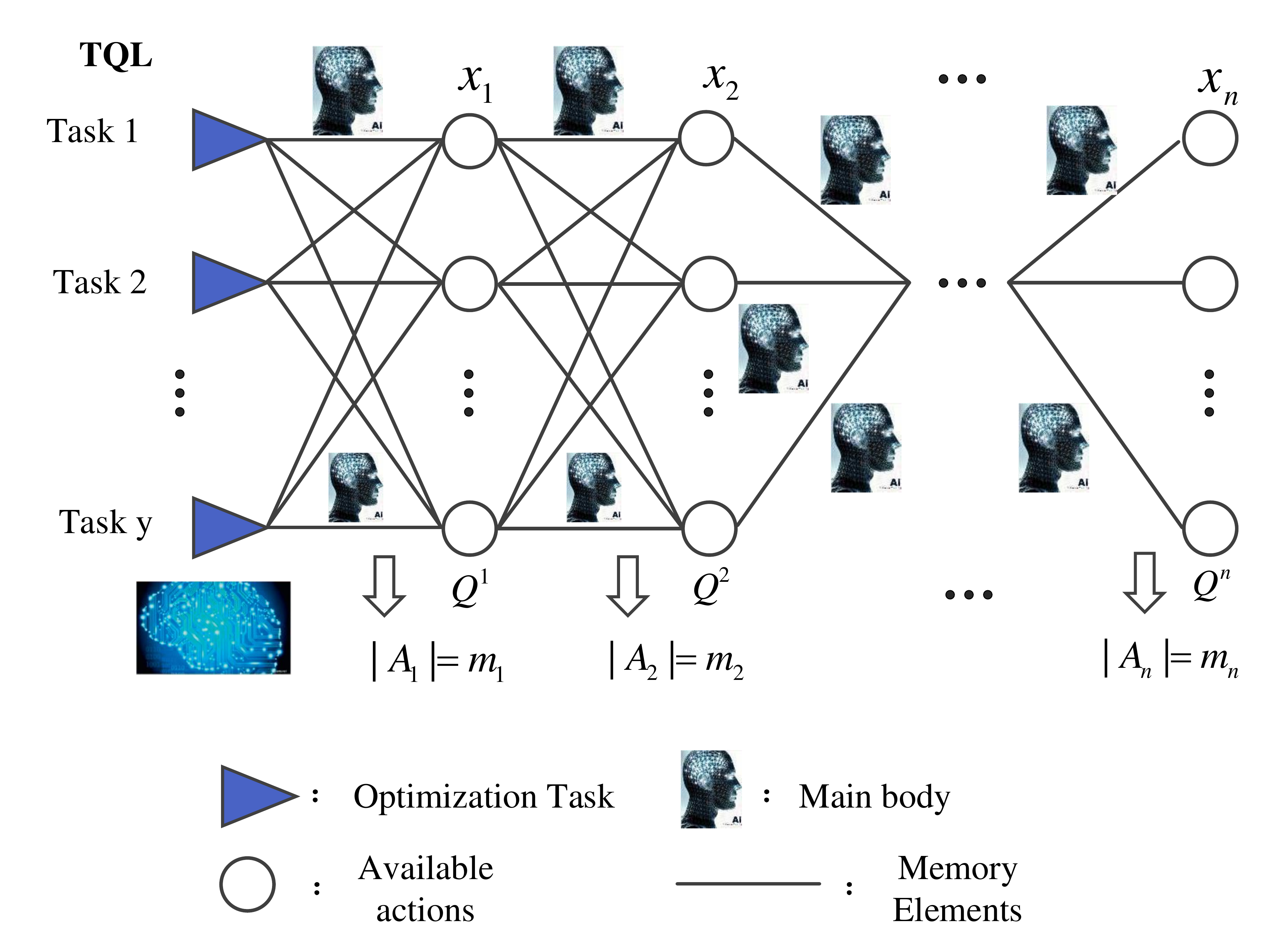}
	\caption{Principle of associative memory}
	\label{f3}
\end{figure}

In order to avoid the curse of dimension, the associative memory is adpoted to reduce state-action space by decomposing the large scale knowledge matrix into multiple lower dimensional space\cite{Zhang2017}. Hence, each controllable variable has a corresponding knowledge matrix. Once the action of the previous variable is determined, this action is taken as the state of the next variable thereby forming a chain connection, as shown in Fig. \ref{f3}. In addition, swarm agent is adopted to improve the knowledge learning rate as there are multiple agents execute actions at the same time and corresponding state-action pairs whose knowledge values are updated simultaneously. After introducing the swarm agent, the $i$th memory matrix ${Q^i}$ can be updated as:
\begin{eqnarray}
\label{Qvalue}
\left\{ \begin{array}{l}
Q_{k + 1}^i\left( {s_k^{ij},a_k^{ij}} \right) = Q_{k + 1}^i\left( {s_k^{ij},a_k^{ij}} \right) + \alpha \Delta Q_k^i\\
\Delta Q_k^i{\rm{ = }}{R^{ij}}\left( {s_{k + 1}^{ij},s_k^{ij},a_k^{ij}} \right) + \gamma \mathop {\max }\limits_{{a^i} \in {A_i}} Q_k^i\left( {s_k^{ij},a_k^{ij}} \right)\\
\quad \quad \;\;\; - Q_k^i\left( {s_k^{ij},a_k^{ij}} \right)
\end{array} \right.
\end{eqnarray} 
{where $\alpha$ is the learning rate; $i$($i$=1,2,...,$n$) denotes the $i$th variable and $j$($j$=1,2,...,$L$) represents the $j$th agents; $n$ and $L$ are the number of variables and agents, respectively;$\gamma$ is the discount factor; subscript $k$ denotes the $k$th iteration; $\Delta Q$ is the knowledge increment;($s_k$,$a_k$) denotes the state-action pair at the $k$th iteration; $R\left( {s_{k + 1}^{},s_k^{},a_k^{}} \right)$ is the feedback reward of a transition from state $s_k$ to $s_k+1$ after exceuted action $a_k$.}

Reinforcement learning methods often adopt a pure strategy of greedy actions or a random global serach strategy. In general, local search based on greedy strategy tends to cause the algorithm to fall into a local optimal solution, while random global search strategy tends to result in a long optimization time. Therefore, this paper uses the $\epsilon$-greedy strategy\cite{re2018} to effectively balance the local search and the global search, as follows:

\begin{eqnarray}
\label{action}
a_{k + 1}^{ij} = \left\{ {\begin{array}{*{20}{c}}
	{\mathop {\arg \max }\limits_{{a^i} \in {A_i}} Q_k^i\left( {s_{k + 1}^{ij},{a^i}} \right),\;{\rm{if }}\;\;{\rm{ }}\varepsilon  \le {\varepsilon _0}}\\
	{{a_s}\quad \quad \quad \quad \quad \quad \quad \quad \quad\;\;{\rm{Otherwise}}}
	\end{array}} \right.
\end{eqnarray}
{where ${\varepsilon _0}$ is a random number with a probability uniformly distributed in [0, 1]; $\varepsilon $ is the exploitation rate representing the probability of a greedy action (exploitation); ${a_s}$ denotes a random action (global search).}

After agents exceuted their actions, a reward to evaluate corresponding state-action pair will be recieved by each agent. In general, an agent will recieve a larger reward if the exceuted action results in a better solution (smaller fitness function). Hence, the reward rule is designed as follows:

\begin{eqnarray}
\label{reward}
{R^{ij}}\left( {s_{k + 1}^{ij},s_k^{ij},a_k^{ij}} \right) = \left\{ {\begin{array}{*{20}{c}}
	{\frac{W}{{f_j^{k + 1}}},{\rm{   }}\quad \quad \quad \quad {\rm{if}}\;f_j^{k + 1} \le f_j^k}\\
	{0,\quad {\rm{   }}\quad\quad\quad\quad{\rm{otherwise}}}
	\end{array}} \right.
\end{eqnarray}
{where $f_j^k$ is the fitness function of the $j$th agent after the $k$th iteration; $W$ is a positive multiplicator.}

\subsection{Learning Efficiency Improvement via Imitation Learning}
In a new environment, agents need to execute a series of random exploitation and exploration process to obtain the optimal policy, and this will consume a quite long time without any prior knowledge.

Thus, the imitation learning is adopted in this section to accelerate the random search process during the initial search process. In the imitation process, RL agents can be regarded as students, which can learn and imitate to other teachers with more knowledge, as shown in Fig.\ref{f2}. In order to guide the RL agents to update the knowledge matrix during the initial phase better, a high efficient whale optimization algorithm (WOA) is adopted as the teacher. In order to avoid agents being trapped in local optimal solution, half of agents learn from WOA to select state-action pair and update the knowledge matrix, the other agents take actions and update based on $\epsilon$-greedy rule shown in (\ref{action}). After each iteration step is completed,  the rewards of all agents are calculated and sorted, and the corresponding state-action pair with largest reward is transmitted to imitative teachers. In a new iteration, the agents with larger rewards excetue actions through e-greedy, while the agents with smaller reward learn from WOA to select state-action pair.  

\subsection{Knowledge Transfer via Transfer Learning}

As illustrated in Fig. \ref{f2}, ITQ agents obtain optimal knowledge matrices for source tasks during the pre-learning process, the prior knowledge are then exploited as the initial knowledge matrices of a new task, thereby avoiding agents' blind explorations and improving search efficiency. This transfer process are designed as:
\begin{eqnarray}
\label{trans}
Q_{_{ni}}^0 = \sum\limits_{h = 1}^H {{r_h}Q_{ni}^*} ,\quad \quad i = 1,2, \cdots ,N 
\end{eqnarray}
{where $Q_{_{ni}}^0$ denotes the initial knowledge matrix of the $i$th variable in the new task; ${Q_{ni}^*}$ represent the optimal knowledge matrix of the $i$th variable in the $h$th source task; $r_h$ represents the similarity between the new task and the $h$th source task; H denotes the number of the source task.}  

\subsection{Convergence analysis}
Let the probability that the number of optimal solutions in the agents is i at time t is ${P_i}\left( t \right) = P\left( {F\left( {s\left( t \right)} \right) = i} \right)$. According to Bayesian conditions Probability formula,
\begin{eqnarray}
\label{c1}
\begin{array}{l}
{P_0}\left( {t + 1} \right) = P\left( {{F_{t + 1}} = 0|{F_t} = 0} \right) \times P\left( {{F_t} = 0} \right) + \\
\quad \quad \quad \quad \;P\left( {{F_{t + 1}} = 0|{F_t} \ne 0} \right) \times P\left( {{F_t} \ne 0} \right)
\end{array}
\end{eqnarray}
{where $s(t)$denotes the state of agents at time $t$ and ${F_{t + 1}} = F\left( {s\left( {t + 1} \right)} \right)$ denotes the number of optimal agents at time $t$.}

Note that the algorithm keeps the information of the optimal agent in each iteration. Therefore, when there is an optimal agent at time $t$, the probability that there is no optimal agent at time $t+1$ is 0, that means $P\left( {{F_{t + 1}} = 0|{F_t} \ne 0} \right) \times P\left( {{F_t} \ne 0} \right) = 0$. Therefore, the probability that there is no optimal agent at time $t+1$ is:
\begin{eqnarray}
\label{c2}
{P_0}\left( {t + 1} \right) = P\left( {{F_{t + 1}} = 0|{F_t} = 0} \right) \times {P_0}\left( t \right)
\end{eqnarray}

In addition, note that the algorithm can search the optimal solution at any step, assuming that the optimal number of agents is zero at time $t$, the probability of searching for the optimal agent at time $t+1$ is not zero, that means  $P\left( {{F_{t + 1}} > 0|{F_t} > 0} \right) > 0$.

With increasing iteration step, let $\tau  = \min \left( {P\left( {{F_{t + 1}} > 0|{F_t} > 0} \right) > 0,t = 0,1, \cdots } \right)$, we can obtain $P\left( {{F_{t + 1}} > 0|{F_t} = 0} \right) \ge \tau  > 0$

When there is no optimal agent at time t, the probability that there is still no optimal agent at t+1 is:
\begin{eqnarray}
\label{c3}
\begin{array}{l}
P\left( {{F_{t + 1}} = 0|{F_t} = 0} \right) = 1 - P\left( {{F_{t + 1}} \ne 0|{F_t} \ne 0} \right)\\
\quad \quad \quad  \;\;\, = 1 - P\left( {{F_{t + 1}} > 0|{F_t} \ne 0} \right) \le 1 - \tau  < 1
\end{array}
\end{eqnarray}

Therefore, when there is no optimal agent at time t, the probability that there is still no optimal agent at t+1 is less than 1, then:
\begin{eqnarray}
\label{c4}
{P_0}\left( {t + 1} \right) \le \left( {1 - \tau } \right){P_0}\left( t \right) \le  \cdots  \le {\left( {1 - \tau } \right)^{t + 1}}{P_0}\left( 0 \right)
\end{eqnarray}

That means
\begin{eqnarray}
\label{c5}
\mathop {\lim }\limits_{t \to \infty } P\left( {{F_{t + 1}} > 0} \right) = 1 - \mathop {\lim }\limits_{t \to \infty } P\left( {{F_{t + 1}} = 0} \right) = 1
\end{eqnarray}

This shows the algorithm can converge to the global optimal solution with probability 1, and the global convergence of the algorithm is proved.
\section{Design of ITQ for Load Parameters Identification}
In this section, the detail steps of ITQ based load parameters identification are introduced and the overall procedure is presented.
\subsection{Design of action and state}
Based on the results in \cite{He2006,Regulski2015}, typical range of each load parameters are shown in Tabel \ref{t1}.

\begin{table}[htb]
	
	\caption{Numerical interval of load parameters}
	\label{t1}
	\renewcommand{\arraystretch}{2}
	\begin{tabular}{cccccccc}
		\hline
		Parameter   & $R_s$  & $X_s$  & $X_m$  & $R_r$  & $X_r$ \\ \hline
		
		Range & [0.02,0.2] & [0.1,0.2] & [2,3.8] & [0.01,0.1] & [0.5,1.5]  \\ \hline
		
		Parameter   & $K_{pm}$  & $H$  & $A$ & $B$ & $a_p$  \\ \hline
		
		Range & [0.2,0.9] & [0.5,2] & [0.2,1] & [0,1] & [0.1,0.9] \\ \hline
		
		Parameter   & $b_p$  & $a_q$  & $b_q$      \\ \hline
		Range & [0.1,0.9] & [0.1,9] & [0.1,9]      \\ \hline
	\end{tabular}
\end{table}

In general, the standard Q learning algorithm is based on discrete Markov processes, and it cannot be directly applied to the solution of continuous variables optimization problems. The discretization method is the most direct means to solve this problem at present. Divide the continuous variables into several parts can approximate the optimal solution of the original problem with sufficient accuracy. In this paper, the continous intervals of each parameter are divided into 100 parts. Then, the action of each variable (load model parameter) is defined by:
\begin{eqnarray}
\label{actionset}
{{\bf{A}}_i} = \left\{ {\begin{array}{*{20}{c}}
	{{a_{i,1}}}&{{a_{i,2}}}& \cdots &{{a_{i,50}}}
	\end{array}} \right\}
\end{eqnarray} 
{where ${\bf{A}}_i$ denotes the $i$th variable's action set; $a_{i,k}(k=1,2,...,50)$ denotes the $k$th action of the $i$th variable.}

As stated in section IV, the action set of each variable is the state set of the next variable, i.e. ${\bf{A}}_i={\bf{S}}_{i+1}$. For the first variable, the state set is the action set.

\subsection{Design of Reward function}
According to the mathematical model described in \ref{obj}, the fitness function  minimize the loss function can be obtained as :

\begin{eqnarray}
\label{reward1}
{R^{ij}}\left( {s_{k + 1}^{ij},s_k^{ij},a_k^{ij}} \right) = \left\{ {\begin{array}{*{20}{c}}
	{\frac{W}{{f_j^{k + 1}}},{\rm{   }}\quad \quad \quad \quad {\rm{if}}\;f_j^{k + 1} \le f_j^k}\\
	{0,\quad {\rm{   }}\quad\quad\quad\quad{\rm{otherwise}}}
	\end{array}} \right.
\end{eqnarray}
Hence, the reward function of each agent can be obtained by \ref{reward} after each iteration step.
\subsection{Design of knowledge transfer}
The key to determine the transfer quality is the definition of the similarity between source task and new task. From \ref{obj} we can see that the optimization task of load parameters identification is determined by the bus voltage, active and reactive power. Hence, Fréchet distance\cite{Eiter94} is adopted to measure the similarity between bou voltage curves,active power curves and reactive power curves in the source tasks and in the new task. The Fréchet distance is the shortest length of the leash (association) allowing them (the determinant and disease) to be on two separate curves from start to finish, which takes into account the location and ordering of the points along the curves and are widely used in curve similarity analysis. Let ${\bf{F}}$ and ${\bf{G}}$ are bus voltage curves in source task and new task, and the length for each cure are ${\bf{T}}$ and ${\bf{W}}$. The bus voltage in the source task as a function of time is given by ${\bf{F}}\left( {\alpha \left( t \right)} \right)$ and ${\bf{G}}\left( {\beta \left( t \right)} \right)$, where $\alpha \left( t \right)$ and $\beta \left( t \right)$ are two increasing functions and $\alpha \left( 0 \right) = 0$,$\alpha \left( 1 \right) = T$,$\beta \left( 0 \right) = 0$,$\beta \left( 1 \right) = W$. Mathematically, the Fréchet distance between two curves is defined as    
\begin{eqnarray}
\label{fdistance}
{\delta _F}\left( {F,G} \right) = \mathop {\inf }\limits_{\alpha ,\beta } \mathop {\max }\limits_{t \in \left[ {0,1} \right]} \left\{ {d\left( {F\left( {\alpha \left( t \right)} \right),G\left( {\beta \left( t \right)} \right)} \right)} \right\}
\end{eqnarray}
{where $d$ is the Euclidean distance function}.

Hence, the similarity between two curves is determined by the equation:
\begin{eqnarray}
\label{fsimi}
SU\left( {F,G} \right) = 1 - \frac{{\mathop {\inf }\limits_{\alpha ,\beta } \mathop {\max }\limits_{t \in \left[ {0,1} \right]} \left\{ {d\left( {F\left( {\alpha \left( t \right)} \right),G\left( {\beta \left( t \right)} \right)} \right)} \right\}}}{{\mathop {\sup }\limits_{\alpha ,\beta } \mathop {\max }\limits_{t \in \left[ {0,1} \right]} \left\{ {d\left( {F\left( {\alpha \left( t \right)} \right),G\left( {\beta \left( t \right)} \right)} \right)} \right\}}}
\end{eqnarray}
{where $SU\left( {F,G} \right) \in \left[ {0,1} \right]$, A value near 1 indicates more	similarity between the two curves, while a value near 0	indicates less similarity between them. }

Similarly, similarity of active power curves  and reactive curves $SP$ and $SQ$ can be computed according to \ref{fsimi}. 
the The Fréchet distance between ${\bf{A}}$ and ${\bf{B}}$ is defined as the infimum over all reparameterizations

similarity between the source tasks and the new task is defined as :
\begin{eqnarray}
\label{similarity}
r = {\omega _1}SU + {\omega _2}SP + {\omega _3}SQ
\end{eqnarray}

\subsection{Overall Procedure}
\begin{figure}[htb]
	\centering                                   
	\includegraphics[width=8cm]{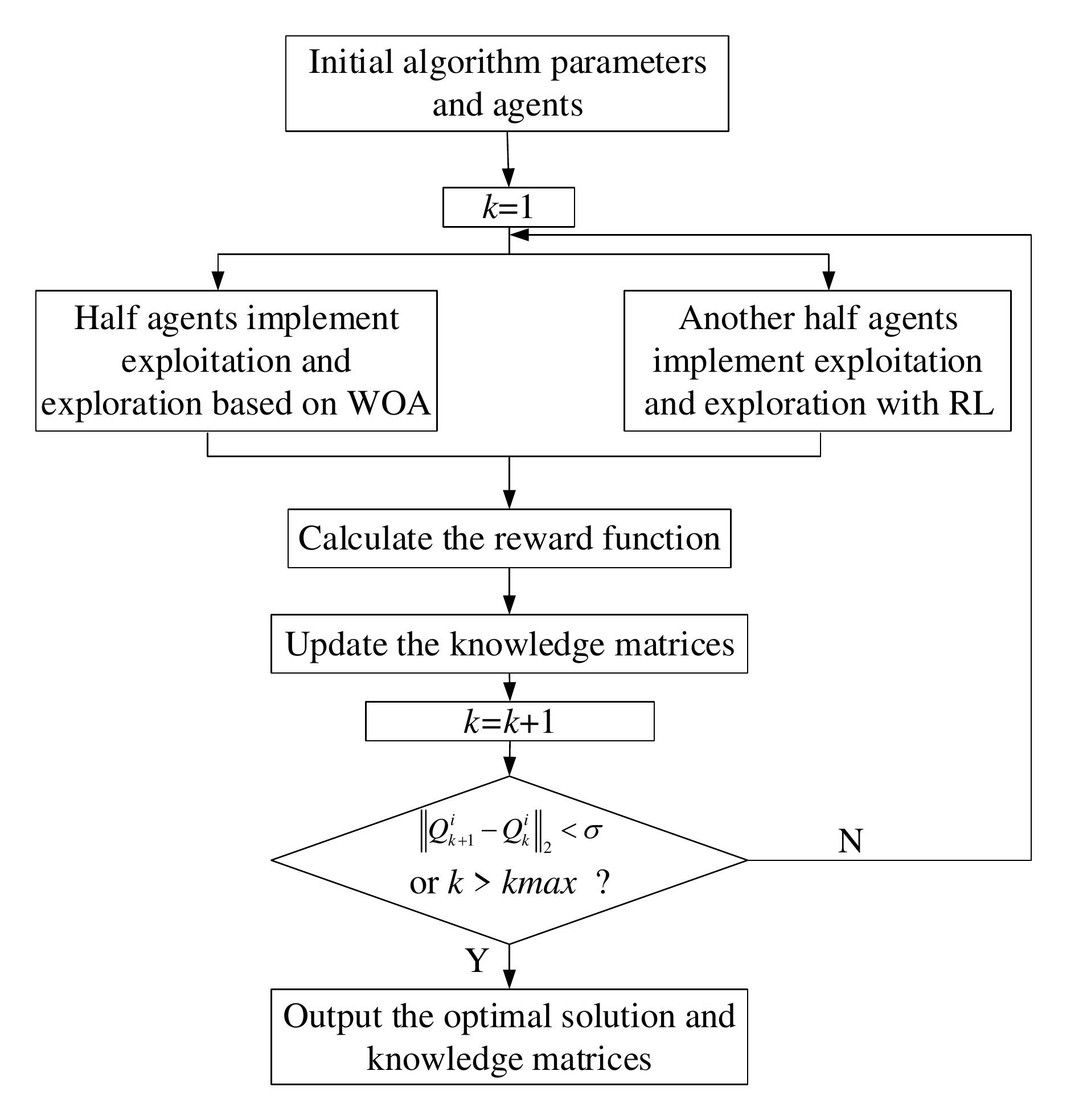}
	\caption{Equivalent circuit of this composite load model}
	\label{f5}
\end{figure}

The overall process to implement the approach is shown in Fig.\ref{f5}, where $k_{max}$ denotes the maximum iteration steps and ${\left\| {Q_i^{k + 1} - Q_i^k} \right\|_2}$ is the Euclidean norm of $Q$-value differences, and $\zeta$ is the convergence coefficient. As shown in Fig.\ref{f5}, the pre-learning process is firstly executed to accumulate the optimal knowledge from the source tasks, then, agents' action strategy in the new task is initialized with transfer learning, thereby accelerating the optimiazation process. In real power systems, the source tasks can be obtained by adopting the past measured fault data.

\section{Case Study}
This section evaluates the effectiveness of the proposed ITQ approach. The estimated results from ITQ are compared with that of the whale optimization algorithm (WOA)\cite{Mirjalili2016}, grey wolf optimizer (GWO)\cite{Mirjalili2014}, IPSO\cite{Regulski2015}, and classical Levenberg-Marquardt(L-M) method\cite{Kim2016}. In order to generate the fault data, simulations are conducted on the New England 68-bus test system with composite ZIP and IM loads. All simulations are undertaken in Matlab Power System Tool (PST) and the sampling rate is 100Hz. The population size and the maximum iteration step are set as 30 for each heuristic algorithms. For ITQ, the parameters are chosen by a few experiments, which are shown in Table \ref{t2}.

\begin{table}[]
	\caption{Parameters used in ITQ}
	\centering
	\label{t2}
	\renewcommand{\arraystretch}{1.5}
	\begin{tabular}{p{2.5cm}<{\centering}p{2.5cm}<{\centering}p{2.5cm}<{\centering}}
		\hline
		Parameter & Pre-learning & Transfer learning \\ \hline
		$\alpha$         & 0.1          & 0.1                 \\ 
		$\gamma$         & 0.2          & 0.1                 \\ 
		$\epsilon$         & 0.5          & 0.8                 \\ 
		$W$         & 1            & 1                   \\ \hline
	\end{tabular}
\end{table}

\subsection{Simulation model}
The 68-bus test systemis a reduced-order model of the New England/New York interconnected system\cite{Cui2019}. It contains 16 generators, 68 buses, and 29 loads. Each load is described as a composite load with ZIP and IM and load parameters identification process is carried out for load connected on bus 27.

\subsection{Pre-Learning Process}
A pre-learning process needs to be firstly executed to accumulate the optimal knowledge matrices from the source tasks for ITQ algorithm. Therefore, 5 different scenarios are simulated and scenario 1 and 2 are taken as source tasks. True load parameters and fault types in each scenario are shown in Table \ref{t3}. For fault type, type $0$ represents three phase fault, type $1$ represents line to ground fault and type $2$ represents line to line fault.   

\begin{figure}[htb]   
	\centering                                
	\includegraphics[width=8.2cm]{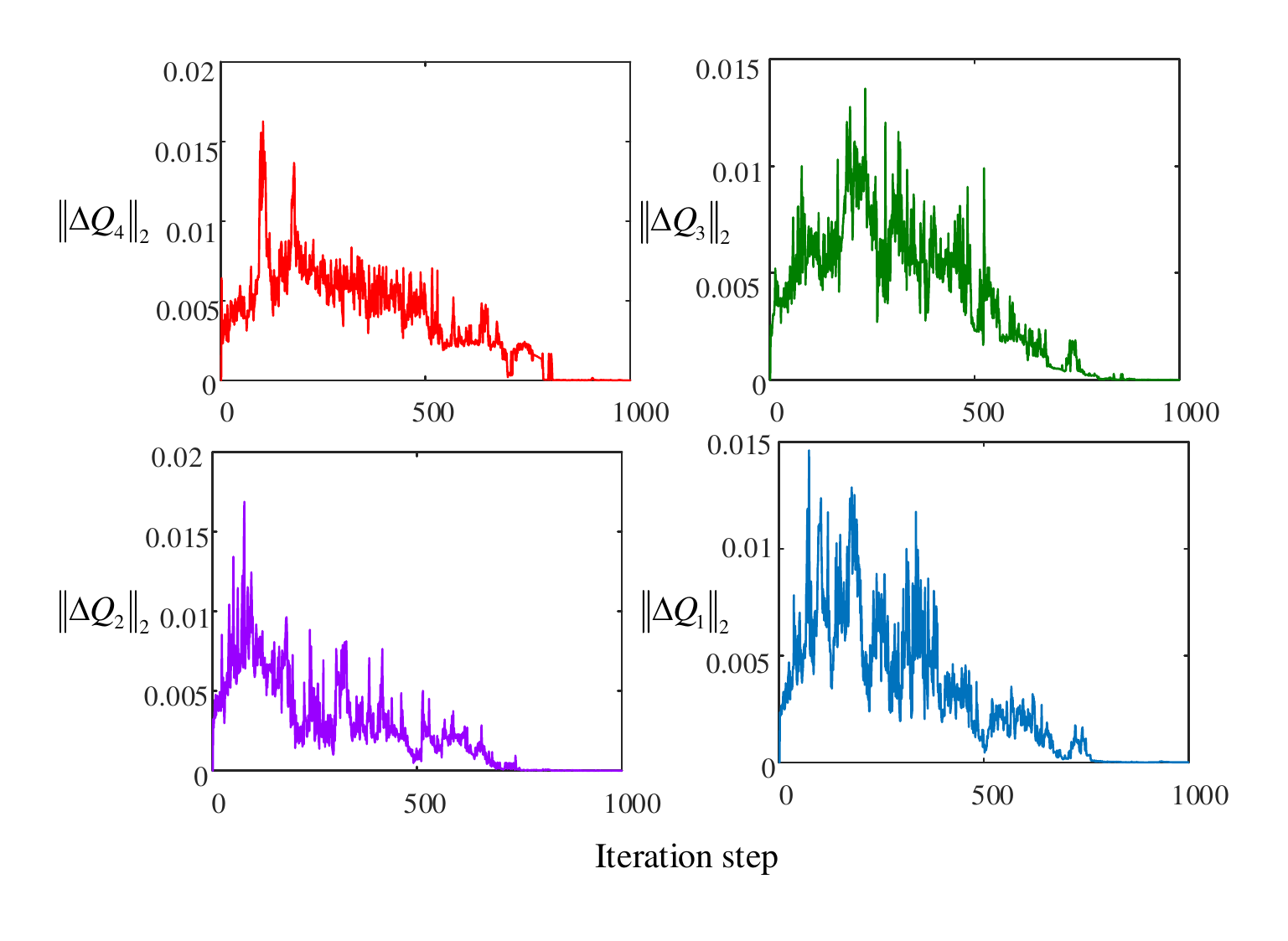}
	\caption{Convergence of the memory matrices}
	\label{f9}
\end{figure}

\begin{figure}[htb]   
	\centering                                
	\includegraphics[width=8cm]{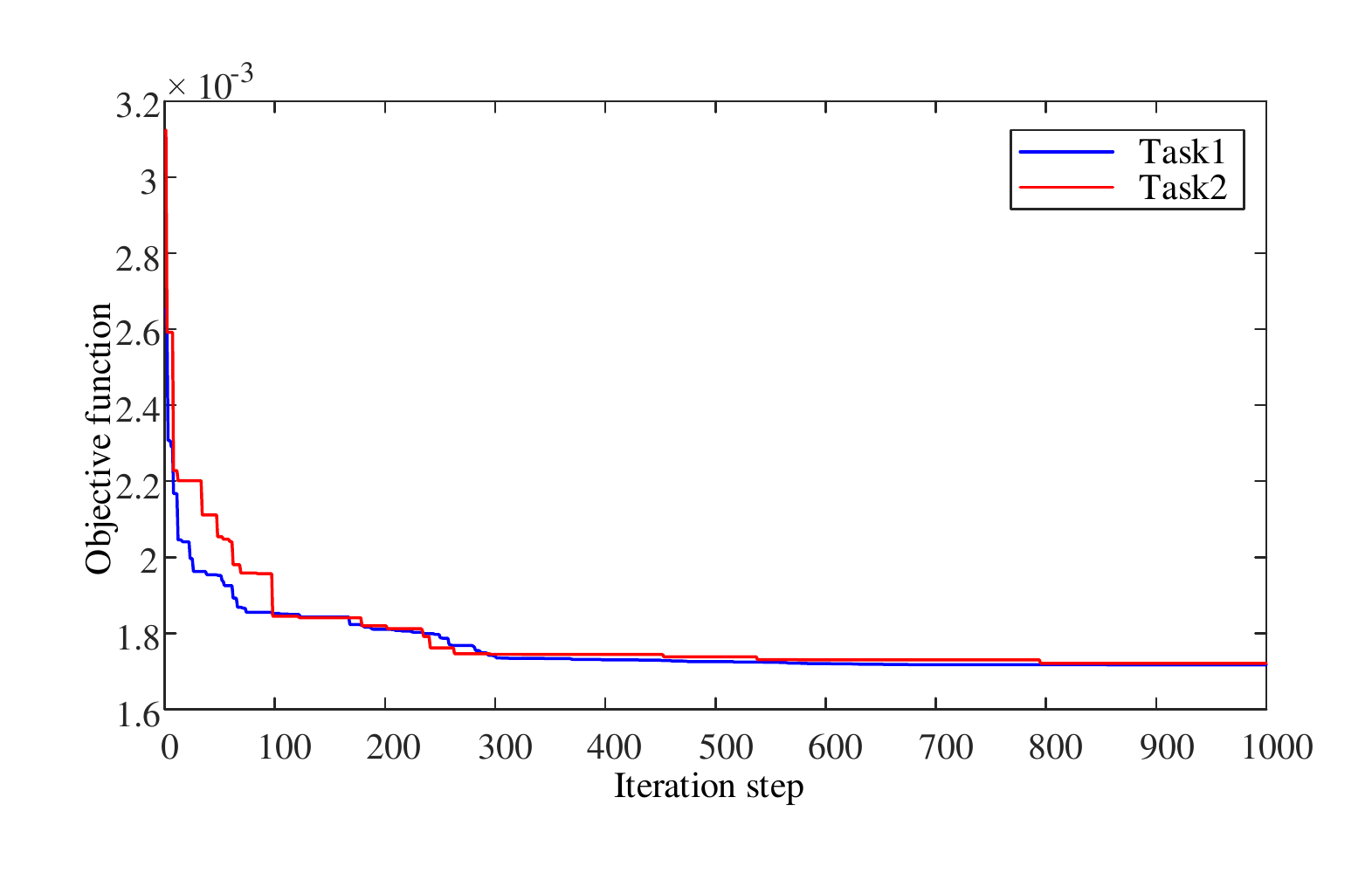}
	\caption{Convergence of the objective functions}
	\label{f14}
\end{figure}

\begin{figure}[htb]   
	\centering                                
	\includegraphics[width=8.2cm]{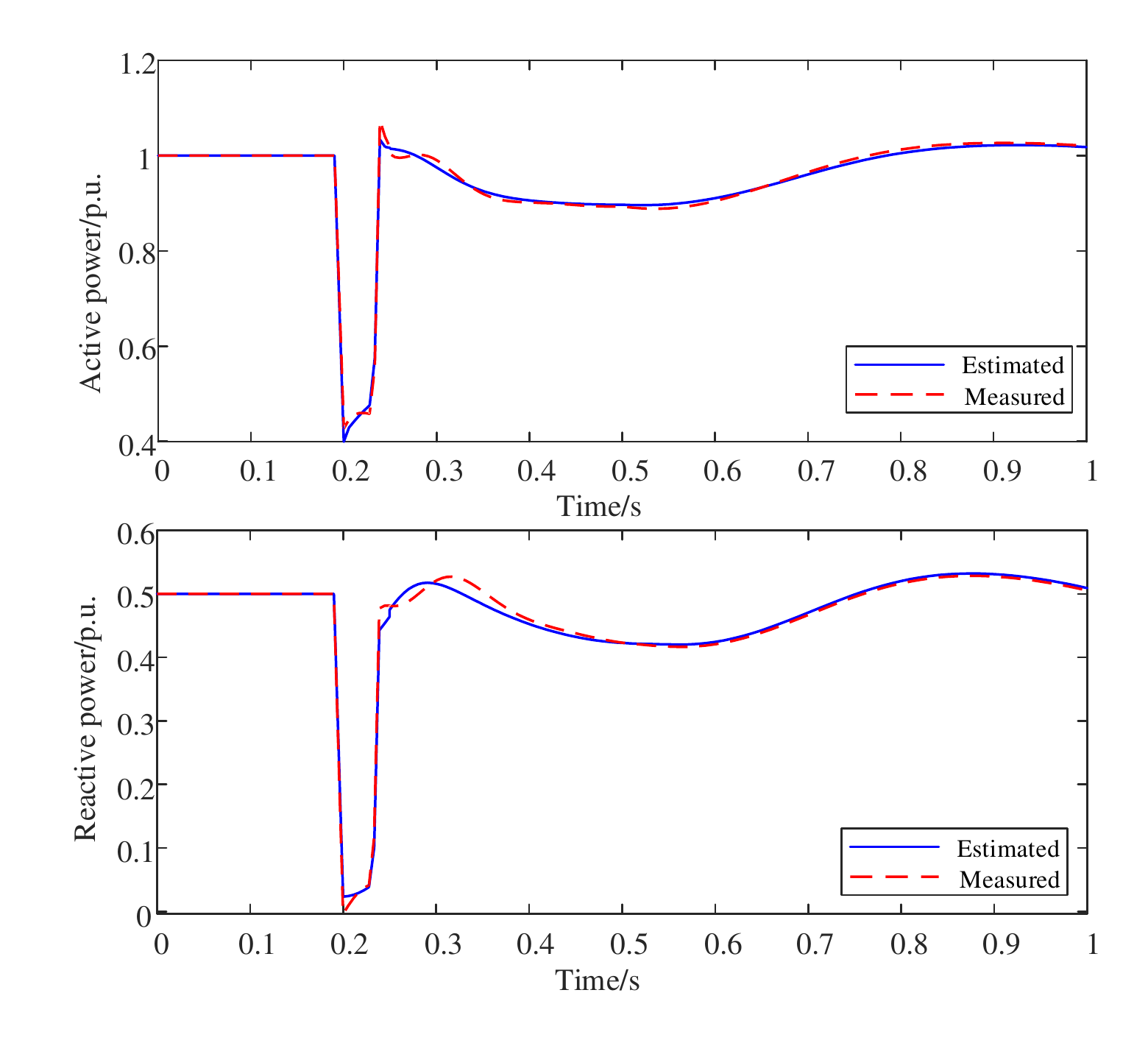}
	\caption{Pre-set parameters for different scenarios}
	\label{f6}
\end{figure}

Fig. \ref{f9} - Fig. \ref{f14} shows the convergence curve with respect to iteration number for scenario 1. It is clear that each variable can converge to its own optimal knowledge matrix after 700 iteration steps. The reward obtained by an agent during the learning process is shown in Fig.\ref{f13} and the optimal fitness function during the learning process among all agents is shown in Fig.\ref{f14}.  It is clear that ITQ can converge to the optimal knowledge matrices for source task 1. Similarily, when applying the pre-learning process on task 2, a high quality fitness function can be obtained, as shown in Fig. \ref{f14}. Fig.\ref{f6} presents the comparison between the estimated power outputs and measurements. It can be seen that the estimated outputs are very close to measurements. These results validate the high convergence stability of the proposed ITQ method in the pre-learning process.

\begin{table*}[htb]
	\centering
	\caption{Frequencies and damping ratios based on Eigen-analysis}
	\label{t3}
	\renewcommand{\arraystretch}{1.8}
	\begin{tabular}{p{0.8cm}<{\centering}p{0.8cm}<{\centering}p{0.8cm}<{\centering}p{0.8cm}<{\centering}p{0.8cm}<{\centering}p{0.8cm}<{\centering}p{0.8cm}<{\centering}p{0.8cm}<{\centering}p{0.8cm}<{\centering}p{0.8cm}<{\centering}p{0.8cm}<{\centering}p{0.8cm}<{\centering}p{0.8cm}<{\centering}p{0.8cm}<{\centering}p{0.8cm}<{\centering}}
		\hline
		Scenario & $R_s$  & $X_s$  & $X_m$  & $X_r$   & $K_{pm}$  & $R_r$   & $a_p$   & $a_q$   & $b_p$   & $b_q$  & $H$  & $A$  & $B$  & fault type \\ \hline
		S1       & 0.045  & 0.173  & 2.49   & 0.131   & 0.43      & 0.031   & 0.40    & 0.30    & 0.30    & 0.30   & 1.2  & 0.9  & 0.10  & 1        \\
		S2       & 0.113  & 0.104  & 2.21   & 0.081   & 0.71      & 0.045   & 0.55    & 0.25    & 0.15    & 0.35   & 1.1  & 0.5  & 0.83  & 0        \\
		S3       & 0.188  & 0.145  & 3.35   & 0.151   & 0.55      & 0.065   & 0.30    & 0.20    & 0.40    & 0.40   & 0.7  & 0.9  & 0.51  & 2        \\
		S4       & 0.151  & 0.112  & 2.83   & 0.163   & 0.62      & 0.021   & 0.61    & 0.15    & 0.23    & 0.42   & 1.4  & 0.7  & 0.29  & 0          \\
		S5       & 0.072  & 0.152  & 3.22   & 0.097   & 0.33      & 0.071   & 0.33    & 0.27    & 0.57    & 0.31   & 0.9  & 0.3  & 0.90  & 1          \\ \hline
	\end{tabular}
\end{table*}

\subsection{Transfer learning and comparison}

\begin{figure}[htb]   
	\centering                                
	\includegraphics[width=8cm]{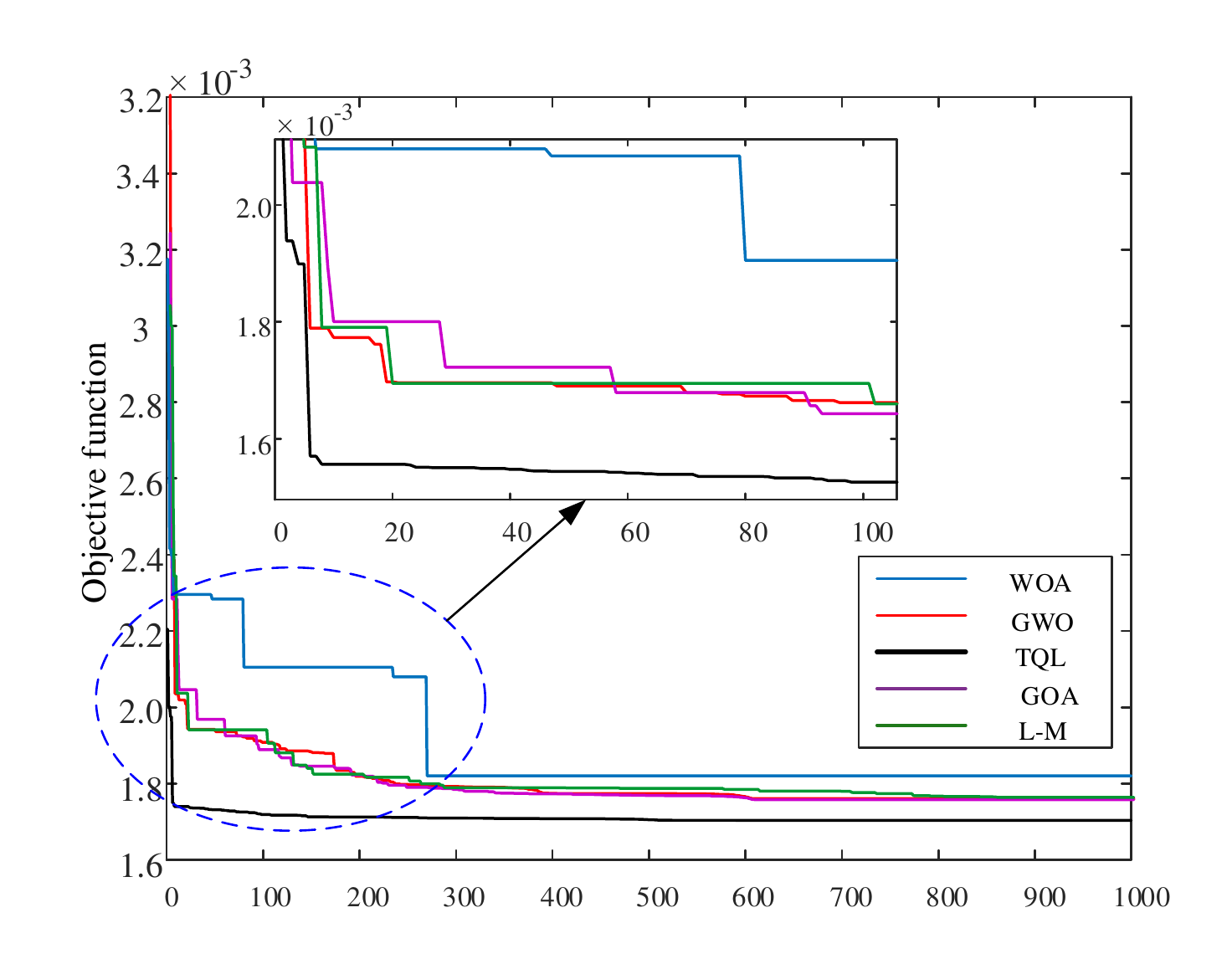}
	\caption{Objective function obtained by five methods}
	\label{f8}
\end{figure}

\begin{figure}[htb]   
	\centering                                
	\includegraphics[width=8cm]{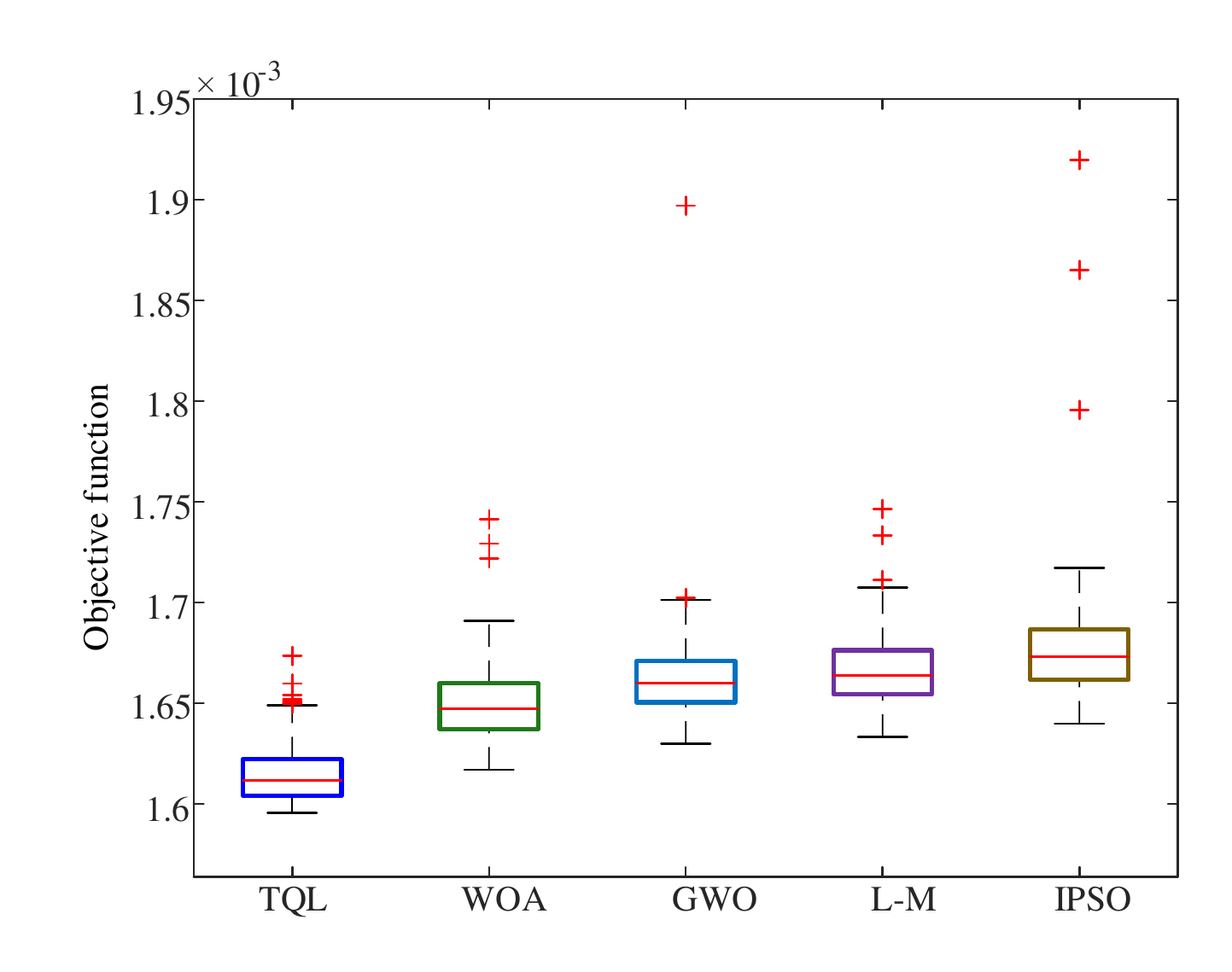}
	\caption{COmparison of Box plot of objective function}
	\label{f7}
\end{figure}

With the pre-learning process completed, the optimal knowledge matrices are exploited for the online load parameters identification tasks by transfer learning. The online identification is implemented for scenario 3.    Based on the definition of similarity in \eqref{similarity},$r_{13}$=0.63 and $r_{23}$=0.71. Therefore, knowledge matrix obtained in task 2 are adopted. Fig. \ref{f8} compares the convergence of the objective function for scenario 3 obtained by ITQ and other 4 algrorithms, including WOA, GWO, IPSO and L-M. From Fig. \ref{f8}, it is clear that ITQ can make a deep exploitation from source tasks when dealing with new task and it can obtain the optimal solution within 150 iteration steps, which is much faster than that of the pre-learning process. The comparison verfies that the convergence rate can be dramatically accelerated by transfer learning. In addition, ITQ can obtain a higher quality reward contributed to the fact that random search agents can avoid the premature convergence and search the globe optimal result. In order to further test the perormance of ITQ, all the algorithms are executed with 100 runs. Fig. \ref{f7} shows the Box plots of objective function obtained by 5 algorithms, and it is clear that ITQ performes best and the convergence stability is higher than other algorithms. .

\subsection{Impact of low similarity and limited source tasks }
This section validate the effectiveness of ITQ with low similarity and limited source tasks. In real power systems, limited source tasks can be an obstacle for transfer learning. Under this condition, the new task can be taken as source task and the pre-learning process should be exuctued to obtain an optimal solution. Fig. \ref{f9} and Fig. \ref{f14} have validated that  ITQ can seach a higher qulity optimal solution compared to exsiting methods even though it cannot exploit the optimal knowledge from source tasks. 
\begin{figure}[htb]   
	\centering                                
	\includegraphics[width=8cm]{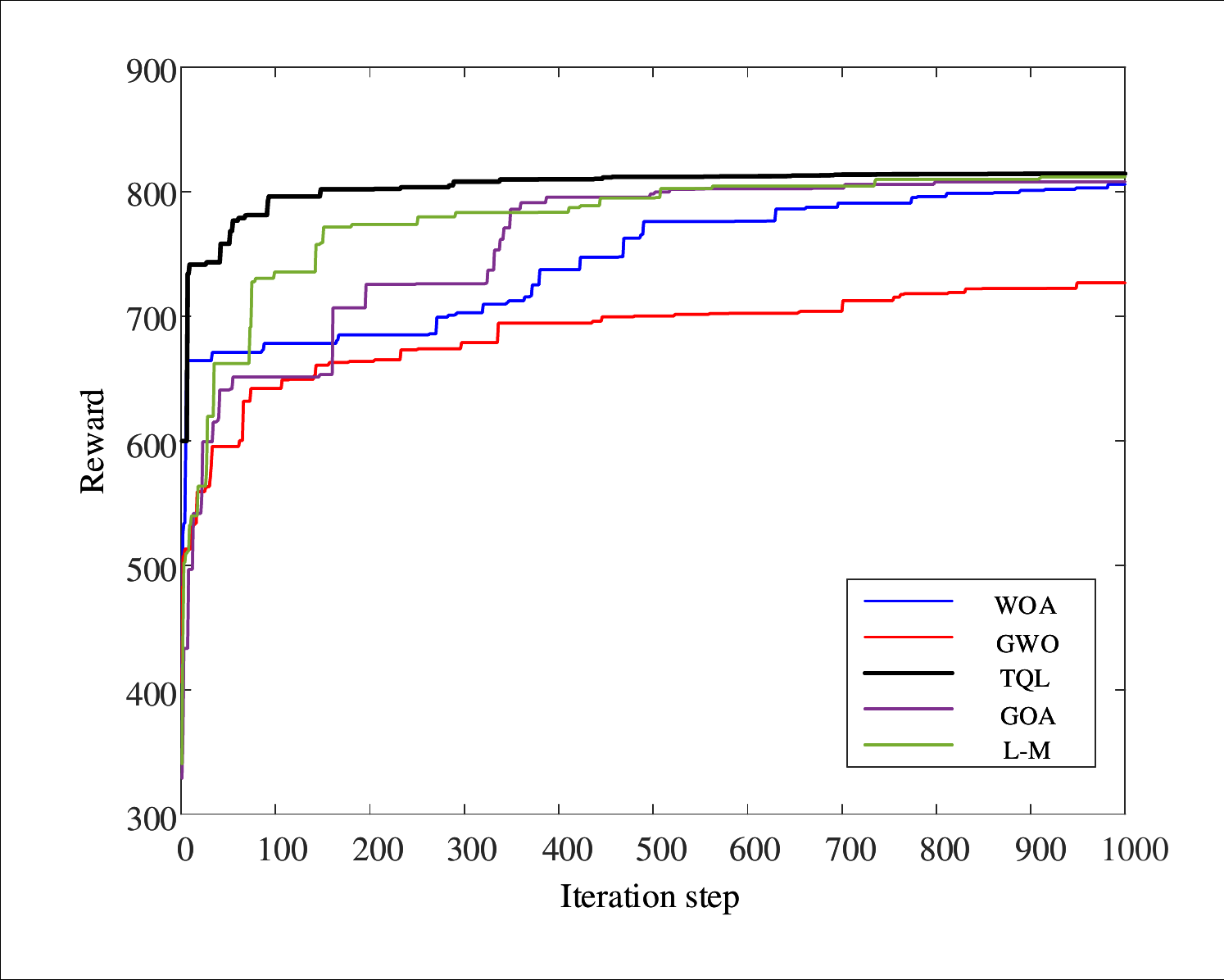}
	\caption{Objective function comparison under low similarity condition}
	\label{f15}
\end{figure}

\begin{table}[]
	\centering
	\caption{Estimated parameters comparison}
	\label{t4}
	\renewcommand{\arraystretch}{1.8}
	\begin{tabular*}{20pc}{@{\extracolsep{\fill}}c|cccccc@{}}
		\hline
		\multirow{2}{*}{Parameter} & \multicolumn{6}{c}{Method}      \\ \cline{2-7} 
		& True  & ITQ    & WOA    & GWO    & IPSO   & L-M \\ \hline
		$R_s$  & 0.151 & 0.1583 & 0.1622 & 0.1631 & 0.1628 & 0.1621 \\
		$X_s$  & 0.112 & 0.1255 & 0.1285 & 0.1311 & 0.1325 & 0.1293  \\
		$X_m$   & 2.83   & 2.909 & 3.11   & 3.023  & 3.152  & 3.106 \\
		$X_r$   & 0.163 & 0.1711 & 0.1832 & 0.1921 & 0.1865 & 0.1955  \\ 
		$K_{pm}$ & 0.62  & 0.6531   & 0.5885   & 0.6959   & 0.6941 & 0.5773  \\   
		$R_r$    & 0.021  & 0.0358   & 0.0322   & 0.0395   & 0.0388 & 0.0331  \\
		$a_p$    & 0.61   & 0.5606   & 0.6963   & 0.5112   & 0.5232 & 0.6885  \\
		$a_q$   & 0.15   & 0.1889   &  0.2213  & 0.2515   & 0.2332 & 0.2106  \\
		$b_p$    & 0.23   & 0.2939   & 0.3121   & 0.2882   & 0.3351 & 0.3025   \\
		$b_q$   & 0.42    & 0.3101   & 0.2859   & 0.1865   & 0.2688 &0.3232   \\	
		$H$   & 1.4   & 1.611   & 1.052   & 1.857   & 1.212 & 1.0886   \\
		$A$   & 0.7   & 0.7414   & 0.7818   & 0.6543  & 0.6852 & 0.7665   \\
		$B$   & 0.29   & 0.3232   & 0.2516   & 0.3568  & 0.2158 & 0.3312   \\\hline
	\end{tabular*}
\end{table}

Similarity analysis shows that $r_{14}=0.51$ and $r_{24}=0.33$, and this indicates that there are few similarities between scenario 4 and another 2 source tasks. ITQ are adopted for scenario 4 to test the performance of with low task similarity. Fig. \ref{f15} shows the comparison optimization results obtained by 5 methods, it indicates that each algorithm can obtain a satisfied results and ITQ presents the smallest objective function which means ITQ has high performance when the similarity between new task and source task is low.

Table \ref{t4} presents identified results from different algorithm for load parameters in scenario 4. For each algorithm, 100trails have been run to obtain the optimal load parameters. From the comparison results shown in Table \ref{t4}, it is clear that ITQ based load parameters are closer to actual values and this is consistent to thre results in Fig.\ref{f14}. There are small discrepancies between estimated parameters and true values, and this may caused by the limited observability of some parametres.

\section{Conclusion and future work}
This paper proposed a transfer reinforcement learning based composite load parameters identification approach to accelerate the identification rate and improve the identification accuracy. A imitation learning process is introduced to improve the exploitation and exploration process of Q learning. The transfer learning process is introduced to overcome the challenge of low efficiency identification. Owing to the greedy search rule, the proposed TIQ can avoid the premature convergence and search the globe optimal result. Simulations on 68-bus system have validated the effectiveness of the proposed TIQ method, and the comparisons show TIQ approach has superior convergence properties owing to the ability to exploit optimal knowledge from source tasks.   

Due to development of the complex load models and high frequency of chanes of time varying load parameters, in the future work, we will consider to extend this approach to identify WECC load parameters and time-varying load parameters .


\bibliographystyle{IEEEtran}
\bibliography{IEEEabrv,aaa}


\end{document}